# Low-energy constraints on photoelectron spectra measured from liquid water and aqueous solutions

Sebastian Malerz[1], Florian Trinter[1,2], Uwe Hergenhahn[1,3], Aaron Ghrist[1,4], Hebatallah Ali[1,5], Christophe Nicolas[6], Clara-Magdalena Saak[7], Clemens Richter[1,3], Sebastian Hartweg[6], Laurent Nahon[6], Chin Lee[1,8,10], Claudia Goy[9], Daniel M. Neumark[8,10], Gerard Meijer[1], Iain Wilkinson[11]*, Bernd Winter[1]*, and Stephan Thürmer[12]*

[1] Molecular Physics Department, Fritz-Haber-Institut der Max-Planck-Gesellschaft, Faradayweg 4-6, 14195 Berlin, Germany
[2] Institut für Kernphysik, Goethe-Universität, Max-von-Laue-Strasse 1, 60438 Frankfurt am Main, Germany
[3] Leibniz Institute of Surface Engineering (IOM), Department of Functional Surfaces, 04318 Leipzig, Germany
[4] Department of Chemistry, University of Southern California, Los Angeles, CA 90089, USA
[5] Physics Department, Women Faculty for Art, Science and Education, Ain Shams University, Heliopolis, 11757 Cairo, Egypt
[6] Synchrotron SOLEIL, L'Orme des Merisiers, St. Aubin, BP 48, 91192 Gif sur Yvette, France
[7] Department of Physics and Astronomy, Uppsala University, Box 516, SE-751 20 Uppsala, Sweden
[8] Department of Chemistry, University of California, Berkeley, CA 94720 USA
[9] Centre for Molecular Water Science (CMWS), Photon Science, Deutsches Elektronen-Synchrotron (DESY), Notkestraße 85, 22607 Hamburg, Germany
[10] Chemical Sciences Division, Lawrence Berkeley National Laboratory, Berkeley, CA 94720, USA
[11] Department of Locally-Sensitive & Time-Resolved Spectroscopy, Helmholtz-Zentrum Berlin für Materialien und Energie, Hahn-Meitner-Platz 1, 14109 Berlin, Germany
[12] Department of Chemistry, Graduate School of Science, Kyoto University, Kitashirakawa-Oiwakecho, Sakyo-Ku, Kyoto 606-8502, Japan

**ORCID**
SM: 0000-0001-9570-3494
FT: 0000-0002-0891-9180
UH: 0000-0003-3396-4511
AG: 0000-0003-2196-9278
HA: 0000-0001-5037-9889
CS: 0000-0001-7898-0713
CR: 0000-0003-1807-1643
SH: 0000-0002-3053-683X
LN: 0000-0001-9898-5693
CL: 0000-0001-9011-0526
CG: 0000-0001-5771-8564
DMN: 0000-0002-3762-9473
GM: 0000-0001-9669-8340
IW: 0000-0001-9561-5056
BW: 0000-0002-5597-8888
ST: 0000-0002-8146-4573

*Corresponding authors: iain.wilkinson@helmholtz-berlin.de; winter@fhi-berlin.mpg.de; thuermer@kuchem.kyoto-u.ac.jp

**Abstract**

We report on the effects of electron collision and indirect ionization processes, occurring at photoexcitation and electron kinetic energies well below 30 eV on the photoemission spectra of liquid water. We show that the nascent photoelectron spectrum and, hence, the inferred electron binding energy can only be accurately determined if electron energies are large enough that cross sections for quasi-elastic scattering processes, such as vibrational excitation, are negligible. Otherwise, quasi-elastic scattering leads to strong, down-to-few-meV kinetic energy scattering losses from the direct photoelectron features, which manifest in severely distorted intrinsic photoelectron peak shapes. The associated cross-over point from predominant (known) electronically inelastic to quasi-elastic scattering seems to arise at surprisingly large electron kinetic energies, of approximately 10-14 eV. Concomitantly, we present evidence for the onset of indirect, autoionization phenomena (occurring via superexcited states) within a few eV of the primary and secondary ionization thresholds. These processes are inferred to compete with the direct ionization channels and primarily produce low-energy photoelectrons at photon and electron impact excitation energies below ~15 eV. Our results highlight that vibrational inelastic electron scattering processes and neutral photoexcitation and autoionization channels become increasingly important when photon and electron kinetic energies are decreased towards the ionization threshold. Correspondingly, we show that for neat water and aqueous solutions, great care must be taken when quantitatively analyzing photoelectron spectra measured too close to the ionization threshold. Such care is essential for both the accurate determination of solvent and solute ionization energies as well as photoelectron branching ratios and peak magnitudes.

# Introduction

The development of liquid-jet photoelectron or more accurately photoemission† spectroscopy (LJ-PES) represents a milestone for research on the electronic structure of liquid water and aqueous solutions.[1-3] Among the quantities of prime interest are solvent and solute lowest vertical ionization energies (VIEs), which measure the energetic cost to detach an electron under equilibrium conditions and thus chart critical parts of the energy landscape that controls chemical reactivity.[4] Furthermore, core-level binding energies (BEs) are sensitive to covalent bonding interactions[5-7] and solute charge states,[8,9] relative peak intensities reveal stoichiometry[10,11] as well as surface propensity (via the so-called depth-probing technique),[12,13] and resonant signal enhancements can be used, *e.g.*, to increase detection sensitivity.[5,14,15] These applications rely on extracting photoelectron (PE) peak areas and/or kinetic energies by taking into account experimental factors such as ionization and electron collection geometries, detection efficiency, and/or ionization cross sections (CSs), although the latter are typically unknown in the aqueous phase.

One often unconsidered and almost overlooked aspect when analyzing PE spectra from liquid water is that PE peak profiles and centroid positions may be altered due to inelastic scattering. Although electron-scattering-induced changes in peak positions have been observed for water clusters as a function of their size[16] and for the solvated electron in liquid water,[17] the prevalent assumption made in condensed-phase PE spectroscopy is that PE peaks are associated with directly-produced photoelectrons that have escaped the sample entirely unscattered. Furthermore, it is generally assumed that peaks (and thus the respective electron binding energies) can be extracted by subtraction of some quantifiable inelastically scattered electron 'background'. However, this is not necessarily the case, as we show here. Indeed, liquid water may seem a favorable case to neglect this electron scattering issue given that inelastic scattering is dominated by electron-impact-induced excitation, neutral dissociation, and ionization at X-ray photoexcitation energies, where photoelectron kinetic energies (eKE) are many tens of electron volts and larger. In fact, for water – a large-band-gap semiconductor[18-21] – the smallest energy for electronic excitation is approximately 7 eV,[22-24] implying that signals from electronically scattered electrons appear at eKEs well below the original direct PE peak. Hence, under high photon and eKE conditions, the nascent direct PE feature profiles are essentially unaffected by inelastic scattering processes. However, the situation changes drastically when photon energies are significantly smaller, such that the primary photoelectrons have insufficient energy to excite/ionize water. In such cases, in both the gas- and condensed phase, vibrational scattering pervades[25-27] and largely determines the scattering-induced changes to the nascent PE spectra. In this article, we will experimentally demonstrate that quasi-elastic scattering similarly leads to photoelectron kinetic energy losses in the liquid-phase, which unlike in the case of high enough eKEs, for all practical purposes, prevents the measurement of accurate binding energies and peak intensities. Furthermore, at specific excitation energies, we will present evidence for valence autoionization resonances in liquid water. These metastable states appear to be accessed both via photon and electron impact excitation close to the primary and secondary ionization thresholds, respectively, competing with direct ionization processes and yielding low-KE primary and secondary electrons.

In a related context, slow electrons play a crucial role in radiation damage in aqueous systems,[28, 29] and hence, the origins and properties of such electrons are essential for understanding, *e.g.*, radiobiological damage. Recently, nascent PE peak distortions have been observed in the low-eKE PE spectra of the hydrated electron, with the distortions being attributed to electron scattering. With the help of detailed scattering models, the nascent ('genuine') electron distribution curve could be recovered.[17]

Specifically focusing on valence ionization of liquid water, experiments have been performed with table-top lasers, so far covering the 1-60 eV photon energy regime, and with soft-X-ray photons from synchrotron radiation facilities, typically with photon energies not lower than 100 eV[‡]. It is the former, lower energy range, and particularly eKEs approaching just a few eV – where the electronic inelastic scattering channels become energetically unfeasible and vibrational and other quasi-elastic processes dominate – that we address in the present work. An associated expectation is that electron binding energies, and particularly the VIE of the lowest energy ionizing transition of liquid water (corresponding to removal of an electron from the leading $1b_1$, highest occupied molecular orbital, denoted as $VIE_{1b_1(l)}$) are not accurately accessible by experiments when too small a photon energy is utilized. This expectation builds on the following basis: At low eKE values, below the electronic excitation threshold of liquid water, the photoelectrons are expected to predominantly engage in vibrational excitations. Focusing on these lowest eKE cases, vibrational scattering processes will produce inelastically scattered background electron signals – most efficiently via just few-meV (single-scattering-event) energy losses – that partially spectrally overlap with the primary PE peaks. Furthermore, as is well known in the radiation chemistry community, within 10-30 eV of the photon or electron impact ionization thresholds, metastable neutral state absorption – followed by dissociation and/or indirect autoionization – can be expected to compete with direct photoionization.[30-32] Such behavior is specifically observed in gas-phase water,[33] and can be expected to occur in liquid water as well.[30-32] Thus, a quantitative analysis of the direct, primary PE peak parameters is expected to be hampered and potentially prevented as photoexcitation energies, and thus eKEs, are progressively reduced. In line with these expectations, we here demonstrate that in liquid water at eKEs less than 15-20 eV, the nascent, directly-produced PE peaks begin to broaden and become increasingly difficult to isolate as the secondary electron impact ionization threshold and predominantly vibrational scattering regime is approached. Notably, since the inelastic scattering behaviors primarily depend on eKE, they should also be detectable using X-rays to ionize high-binding-energy electrons, provided the specific photon energy is close enough to a core-level ionization threshold. We note that for core-level ionization, additional processes may occur close to the ionization threshold that may further distort the PE peaks (post-collision interaction, PCI,[34-37] effects are an example), where we ignore such effects here as they are negligible in our exemplary systems. Indeed, the core-level ionization spectra analyzed and discussed below, reveal similar electron scattering signatures as those observed in the vicinity of the valence photoionization and primary photoelectron impact ionization thresholds.

The first part of the present work reports on valence photoemission measurements of liquid water using continuously tunable 10-60 eV photon energies (hv), which result in photoelectron production in the 0-50 eV KE range. Such LJ-PES experiments have not been technically realized before, and became possible by implementing the wide-energy range VUV/soft X-ray synchrotron radiation beamlines DESIRS and PLÉIADES (both at SOLEIL, St. Aubin) in conjunction with spectrometers that are capable of accurately and efficiently detecting eKEs from aqueous solutions down to nearly-zero electron volts. These studies include threshold measurements of water's fractional PE spectrum produced with photon energies extending below (the centroid of) the $VIE_{1b1(l)}$ water ionization, *i.e.*, close to the ionization onset, revealing that the spectra still exhibit peak-like shapes at threshold but that these 'peaks' are almost entirely generated by inelastically scattered directly, and likely indirectly, produced electrons. The second part of this work presents soft-X-ray photon energy results (measured at the UE56-2_PGM1 beamline at the BESSY II synchrotron facility, Berlin) from a 3 M NaCl aqueous solution, with a focus on the Cl 2p PE spectrum, obtained for photon energies within a range of ~7-18 eV above the respective core-level ionization threshold of approximately 202 eV.[38] Both sets of experiments aim to determine the lowest photon energy, and hence eKE, that can still be used to reveal accurate (nascent) spectral features, *i.e.*, for which the true (essentially undistorted) PE spectrum can still be extracted. Expressed as eKEs, this minimum energy is found to be approximately 10-14 eV for water, where the PE peak intensities diminish, and widths increase, until towards yet smaller energies, signal from direct photoemission can barely be identified on the large background due to inelastic (including quasi-elastic) scattering, at eKEs approximately <10 eV. The background signals specifically underlying the primary electron peaks can generally be associated with small energy losses typical, for instance, of intra- and intermolecular vibrational excitations. On a related note, processes which indirectly produce low kinetic energy photoelectrons and associated electronic scattering channels will also be argued to come into effect at these low excitation energies, something which may be responsible for a surprisingly large background signal peaking near zero eKE. Such signals are found to be consistent with superexcited state[30-32] population and pre-ionization/autoionization of liquid water, either directly following photoabsorption or through primary photoelectron impact excitation. Our data is analyzed with reference to electron collision cross sections for gas-phase $H_2O$ and the condensed-phase $H_2O$ excitation and ionization literature. Similar scattering behavior is observed from NaCl aqueous solutions, as investigated with soft X-ray radiation and core-level ionization.[39] Crucially, our cumulative results allow us to advise that future attempts to measure accurate solute binding energies and peak intensities, particularly for electron detachment from aqueous-phase anions,[4] should be performed at photon energies that are sufficiently in excess of the ionization threshold, where sufficient is here determined to be ~30 eV.

**Experimental**

Measurements of the liquid water valence band were conducted at the PLÉIADES (hv = 20-60 eV) and DESIRS (hv = 10-25 eV) beamlines[40] of the SOLEIL synchrotron facility, Paris, using the PLÉIADES liquid-jet source[41] and the *EASI* (Electronic structure from Aqueous Solution and Interfaces) liquid-jet PES instrument,[42] respectively. The electron–electron coincidence measurements from the 3 M NaCl solution were carried out at the UE56-2_PGM1 beamline[43] of the BESSY II synchrotron facility, Berlin, using a liquid-jet setup coupled to a magnetic bottle time-of-flight electron analyzer, described elsewhere.[44] For the liquid-water measurements, a small amount (~50 mM) of NaCl salt was added to highly demineralized water (conductivity ~0.2 μS/cm) to maintain electrical conductivity and mitigate potentially deleterious sample charging effects.[45] This is common practice when measuring PE spectra from liquid water.[2] The 3 M concentration solutions were prepared by dissolving NaCl (Sigma-Aldrich, ≥99% purity) in highly demineralized water. For the SOLEIL experiments liquid microjets were generated by injecting liquid water (containing 50 mM NaCl) into the interaction vacuum chamber through 40-μm or 28-μm diameter glass capillaries, at a typical flow rate of 0.8 mL/min. Experiments with the 3 M NaCl aqueous solution at BESSY II used a 30-μm capillary at 1 mL/min flow rate.

**Liquid water PES experiment at PLÉIADES**

The measurements were performed using the electromagnetic HU256 undulator and the low-energy 400 lines/mm grating of the beamline. The energy resolution and photon-beam focal spot size (vertical × horizontal) were approximately 2.5 meV and 50 × 120 μm$^2$, respectively. The energy resolution of the hemispherical electron analyzer (wide-angle lens VG-Scienta R4000) is 50 meV at a pass energy of 50 eV. The electron spectrometer is mounted with the electron detection axis perpendicular to the plane of the electron orbit in the storage ring.[46] The light traveled orthogonally to both the liquid jet and the electron detection axis, both being perpendicular to each other. The light polarization was set parallel to the spectrometer axis. While it is unfortunate that measurements could not be conducted in the so-called magic-angle geometry (an angle of 54.7° between the light polarization vector and electron detection axis), which would cancel all intensity variations from the photoelectron angular anisotropy, this has only a minor impact on the results for the following reasons.

Near-equivalent liquid water valence anisotropy parameters, β < 1, have been measured at photon energies below 60 eV for all open ionization channels (0.51 ± 0.06 for $1b_1^{-1}$, 0.75 ± 0.13 for $3a_1^{-1}$, and 0.46 ± 0.13 for $1b_2^{-1}$ at 35.6 eV[47] and 0.27 ± 0.07 for $1b_1^{-1}$, 0.24 ± 0.09 for $3a_1^{-1}$, and 0.18 ± 0.06 for $1b_2^{-1}$ at 29.5 eV[48]); PE signal intensity scales with I ~ 1 + β $P_2[\cos(\theta)]$, where $P_2[x]$ denotes the second order Legendre polynomial. A parallel alignment (θ = 0°) leads to anisotropy-induced signal variations between the open ionization channels of only $I_{3a_1}/I_{1b_1}$ ~ 1.16 and $I_{1b_2}/I_{1b_1}$ ~ 0.97 at 35.6 eV (instead of a value of 1), *i.e.*, a slight enhancement of the $3a_1$ feature at most, which is however below our error bars. Using the values for 29.5 eV the signal variation reduces to less than ~8% with $I_{3a_1}/I_{1b_1}$ ~ 0.98 and $I_{1b_2}/I_{1b_1}$ ~ 0.93.

The liquid jet source is based on a Microliquids© design. Crucially for the experiments discussed here, the liquid jet formed by a 40-μm orifice diameter glass capillary and collected by a heated copper-beryllium catcher was contained within an approximately $7 \times 8 \times 15$ cm$^3$ aluminum enclosure. This box has two 3-mm diameter holes for the synchrotron light to enter and exit, and one 5-mm hole in a titanium piece through which the emitted electrons pass on their way to the hemispherical electron analyzer. When the liquid jet head is inserted, two channels of 1 mm diameter and 5 mm length face the liquid entrance and exit holes, in addition to a 300-μm stainless-steel skimmer facing the titanium hole. The jet is placed at the working distance of electron analyzer (*i.e.*, at 3.4 cm from the 4 cm diameter entrance aperture), which corresponds to a distance of 1 mm between the entrance of the skimmer and the jet. A spectrometer electron transmission measurement using the liquid jet has not been attempted, and the profile of the true, low-eKE, spectral tail is unknown. However, we take the observed rather constant overall spectral shape upon variation of bias voltage (and hence variation of measured eKE range) as an indication of smooth variations of the transmission function in the eKE range considered here.

The liquid feed to the glass capillary was made of non-conductive PEEK line. A small gold-coated metallic connector located 20 cm up-stream before injection into vacuum was used to electrically ground the jet, or to apply a bias voltage to the sufficiently conductive liquid sample. The jet and catcher were always at the same potential. The bias was applied using a highly stable voltage supply (Delta Electronika, SM 70 – AR 24). The liquid solution was pushed through the system using a HPLC (WATREX P102) pump.

The differential pumping box was evacuated with an 800 L/s turbo-molecular pump (Edwards, STPA803C) with a 100 m$^3$/h dry multistage root backing pump (Adixen, A103P). The spectrometer was pumped with one 600 L/s and one 450 L/s turbo-molecular pumps (Edwards, Seiko Seiki STP 600C and STP 450C) with the beamline 600 m$^3$/h dry multistage root backing pump (Edwards, GX600n). Both the spectrometer and the liquid jet chamber are equipped with a liquid nitrogen trap (8400 L/s pumping speed for water). Pressures of $3 \times 10^{-4}$ mbar and $8 \times 10^{-6}$ mbar were achieved in the differential pumping stage and the spectrometer chamber, respectively.

**Liquid water PES experiment at DESIRS**

Experiments at the VUV variable polarization undulator beamline DESIRS[40] were performed in the 10-25 eV photon energy range with the EASI liquid-jet PES setup,[42] which is equipped with a Scienta-Omicron HiPP3 differentially pumped hemispherical electron analyzer. This device uses a unique pre-lens system optimized for the detection of low-energy electrons. Unlike in the instrument described in the previous paragraphs, for the measurements at DESIRS the liquid jet was not enclosed. However, the EASI instrument's efficient μ-metal shielding and low-energy lens mode enabled detection of low-energy electrons. Here, the approximately 28-μm diameter liquid microjet was positioned at a 0.5-0.8 mm distance from the 800-μm orifice diameter skimmer at the analyzer entrance. Similarly, a systematic measurement of the electron transmission function has not been attempted with the EASI instrument. The synchrotron light propagation direction was orthogonal to the liquid jet, both lying in the horizontal plane. We used the hemispherical electron analyzer positioned at a 40° angle with respect to the photon beam propagation direction, with its lens lying in a vertical plane, and a vertical polarization of the photon beam. Although the hemispherical electron analyzer alignment with respect to the light polarization axis deviated somewhat from an ideal value in these experiments (40° instead of 54.7°, *i.e.*, the single-photon ionization magic angle), associated effects on the measured ionization-channel-resolved photoelectron yields are expected to be negligible for the following reasons. Analogous to the discussion for the PLÉIADES experiment, the near-equivalent and near-zero liquid-phase water β values measured at photon energies below 30 eV[48] give anisotropy-induced signal variations of $I_{3a1}/I_{1b1}$ ~ 0.99 and $I_{1b2}/I_{1b1}$ ~ 0.97 at 29.5 eV for an angle of θ = 40°, *i.e.*, less than 4%. At lower photon energies, the liquid water anisotropy parameters are expected to monotonically converge to zero and isotropic emission behaviors as the ionization thresholds are approached.[47] Accordingly, still lower anisotropy-induced signal variations are expected in our threshold (hν ≤ 25 eV) ionization experiments.

The pressure in the main chamber was kept at approximately $5 \times 10^{-4}$ mbar using two turbo-molecular pumps (with a total pumping speed of ~2600 L/s for water) and three liquid-nitrogen cold traps (with a total pumping speed of ~35000 L/s for water). A similar sample delivery design as described above was used to generate the liquid jet, also allowing for the precise application of a bias voltage. The solution was delivered using a Shimadzu LC-20 AD HPLC pump that incorporates a four-channel valve for quick switching between different solutions. The system was equipped with an in-line degasser (Shimadzu DGU-20A5R). The bias voltage was applied using a highly stable Rohde & Schwarz HMP4030 power supply.

All photoemission measurements reported here were conducted in the 10-25 eV photon energy region, using the 200-lines/mm grating of the DESIRS beamline, and with the monochromator exit slit set to 20 μm. These settings yielded an approximately 200 μm horizontal (in the direction of the liquid jet propagation) and 80 μm vertical focus. The energy resolution is exit-slit-limited and given by $\Delta E [eV] = 1.16 \times 10^{-4} \times E [eV]$ (for instance, about 2.3 meV at 20 eV). The exit-slit limited photon flux amounted to ~$4 \times 10^{11}$ ph/s between 10 and 14 eV, $3 \times 10^{11}$ ph/s at 20 eV, and $8 \times 10^{10}$ ph/s at 25 eV photon energies. The energy resolution of the hemispherical electron analyzer was approximately 30 meV at the implemented pass energy of 5 eV. A few spectra recorded with the same set-up, but using a He gas discharge lamp as a laboratory light source are shown in the Supplemental Information (SI).

**3 M NaCl coincidence experiment at BESSY II**

This experiment was performed with the synchrotron operated in 'single-bunch' mode (1.25 MHz light pulse repetition rate). Electrons were detected using a magnetic bottle time-of-flight (TOF) electron analyzer, optimized for the high background pressure encountered in liquid-jet experiments. Details are described elsewhere.[44] The analyzer was aligned vertically, thus the liquid jet, the synchrotron radiation propagation, and the TOF-axis directions were mutually orthogonal. The polarization vector of the synchrotron radiation was vertical, hence coinciding with the analyzer TOF axis. A small accelerating potential into the analyzer was produced by biasing the tip of the magnet at –2 V, and for some spectra measured at 210 and 212 eV photon energy, an additional positive bias voltage was applied to the entrance diaphragm of the spectrometer (facing the liquid jet). Inside the time-of-flight analyzer, electrons were accelerated by a +2 V potential to produce flight times below the temporal bunch spacing of the storage ring. Event-based data acquisition was carried out with a multi-hit capable time-to-digital converter with a 60 ps bin width (GPTA, Berlin). Flight times were measured against a clock signal providing the revolution frequency of electrons in the BESSY II storage ring, and converted to kinetic energies using measured reference spectra. A coincidence analysis of the set of events recorded within 60 s/photon energy was performed, in which we retained only electron pairs with a fast electron in the eKE range (taken as 150-200 eV) of the Cl LMM Auger decay and a slow electron of any smaller kinetic energy ('coincident electrons'). Alternatively, the undiscriminated electron spectrum can be produced from the full set of events.

## Results and Discussion

**Full photoemission spectra from liquid water: hv ≥ 20 eV**

Figure 1 presents a series of full photoemission spectra from liquid water obtained at photon energies between 20 and 60 eV. 'Full' refers to spectra extending from the kinetic energy of the low-energy cutoff, $E_{cut}$, at the onset of the large signal tail (appearing at KE < 10 eV) generally associated with inelastically scattered electrons, up to the lowest-ionization-energy (highest eKE) valence emission feature, assigned to photoemission from the $1b_1$ molecular orbital. All spectra were recorded from a water jet biased at -55 V; the resulting shift in the measured eKEs has been subtracted in Fig. 1 such that the PES spectra appear as if being measured from a grounded jet. The reason for applying a bias voltage is that $E_{cut}$ can be separated from the corresponding low-energy cutoff arising from the electron analyzer itself.[49] Arguably more important for this study, electron signal contributions from gas-phase water around the liquid jet can be effectively removed from the PES spectra. With the focal size of the photon beam and analyzer being larger than the liquid-jet diameter, gas-phase water molecules will be ionized at different distances from the jet, and thus take up different energies in the electric field between biased jet and grounded electron detector. As a consequence, the gas-phase signal is strongly broadened, and the majority of it is mapped to spectral regions that have no overlap with the liquid features. Hence, after correcting for the bias voltage, a nearly pure PES spectrum from neat liquid water is obtained; the effect is shown in Fig. SI-1, panel A of the Supplementary Information (SI), which presents PES spectra measured with and without applied bias voltage, respectively. Intensities of the spectra in Fig. 1 are displayed as measured, except for intensity corrections to account for small variations in photon flux when changing the photon energy. Note that the intensity maxima of the scattering tails are clipped in Fig. 1, and full-intensity-range spectra are shown in Fig. SI-2 of the SI.

In each tier of Fig. 1, we also present peak fits to the water valence spectrum (in blue), where the contributions from the four valence orbitals ($1b_1$, $3a_1$ doublet, $1b_2$, and $2a_1$, marked in the top tier) are cumulatively fit by five Gaussians. We constrained the two $3a_1$ components to have the same height and width.[50] Additionally, the $1b_2$–$3a_1$ peak separation for hv = 25 eV and the $3a_1$–$1b_1$ separation for hv = 20 eV were constrained to fit these peaks on top of the steeply sloping background. The valence peaks shift to lower eKEs with decreasing photon energy (hv) according to KE = hv – BE, where the binding energies (BEs) pertain to the direct ionization energies, VIE, of the respective water molecular orbitals. For hv >50 eV, all four valence PE peaks are well visible on a rather smooth background, while for hv <35 eV, the valence spectrum resides on top of a background of increasing slope, with the water $2a_1$ peak beginning to become undetectable due to diminishing intensity and overlap with the background signal where the low-energy spectral tail rises steeply. Upon further lowering the photon energy, all other valence peaks sequentially disappear as well, and at hv = 20 eV, the remaining emission due to water $1b_1$ ionization results in only a small shoulder near 10 eV eKE. Taken together, we observe a sudden decline of the primary, direct PE peak intensities (yet to be justified in detail), counter-balanced by the relative contribution of the underlying inelastically scattered background signal – most likely including indirect electron production at lower energies – rising steeply when the photoelectron eKEs drop below ~10-14 eV.

The dashed black curves in Fig. 1 represent the contribution from this background signal. Here, we include all broad features (>~4 eV full-width half-maximum, FWHM) within this background, and the sharp cutoff is modeled with exponentially modified Gaussians. Note that the Shirley or Tougaard algorithms commonly applied for inelastic background determination in solid-state X-ray PE spectroscopy are not applicable here for several reasons. The simple Shirley method is unsuited for condensed matter with a strongly eKE-dependent scattering probability, which is the case for semiconductors and insulators.[51] The Tougaard algorithm, on the other hand, is only applicable under the conditions of using absolute-intensity-calibrated spectra together with the correct scattering function for the material.[52] Furthermore, the Tougaard algorithm applies at sufficiently high eKEs, far away from the scattering tail, since this algorithm is incapable of quantifying impact ionization cascades. Neither condition is fulfilled here. Our simple approach to quantify the magnitude of the background signal, particularly that underlying the direct PE peaks, is sufficient and robust, yielding good agreement with available gas-phase photoionization cross sections (CSs), as we will demonstrate below.

**Kinetic-energy-dependent composition of the low-energy spectral tail**

Before we move on to discuss why photoelectron peaks can be severely distorted in some cases, reflecting the rich ionization and scattering behavior of liquid water, we would like to briefly introduce some terminology describing the overall PE spectral shape, including the background signal extending down to zero eKE. Photoelectrons that lose almost all of their initial energy in various scattering processes will give rise to a low-KE tail, denoted here LET, characteristic for condensed-phase photoemission. It is important to realize that the LET spectrum is generally comprised of primary electrons which have lost energy due to (1) various inelastic scattering processes (inelastically scattered primary electrons), as well as (2) electrons formed in impact-ionization cascades that generate secondary electrons, each having sufficient energy to overcome the surface barrier of the sample; electrons with the smallest energies (quasi-zero kinetic energy) give rise to the steep signal edge at the cutoff. The terms 'secondary electron energy distribution' (SEED) or 'secondary electron emission', typically used to denote the LET in the condensed-matter PES literature[53] as well as in electron microscopy and high-energy physics contexts,[54, 55] are misleading if used to describe low-energy spectra. Specifically, these terms do not account for the contribution to the LET of those direct photoelectrons that have lost energy in processes not involving the generation of another electron. This contribution is indeed sizable in the case of a semiconductor excited at very low photon energies, where the eKE is smaller than the band gap, and hence insufficient to ionize another electron. In such a case, the LET will instead consist to a large extent of the inelastically scattered primary photoelectron distribution, here denoted as IPED. Quantification of the latter is elusive due to our currently incomplete understanding of all contributing scattering processes in liquid water. In fact, there is an ongoing lively discussion about the correct modeling of electron scattering in liquid water, especially for the low-energy regime.[56]

Notably, low-KE electrons may also be produced via indirect primary or secondary ionization processes. In the gas phase, at excitation energies up to a few-tens of eV above valence ionization thresholds, direct ionization is known to compete with metastable superexcited state production, with subsequent dissociation and/or indirect (auto-)ionization.[30-32] The indirect ionization processes occur through the coupling of the aforementioned superexcited states to the ionization continuum, producing electron distributions which can extend from zero eKE to the (coupled photoionized state's) adiabatic ionization threshold, depending on the degrees of internal excitation produced in the residual photoionized species. Such states have been found to play notable roles in the valence ionization dynamics of gas-phase water[57] and amorphous ice[58]. However, the degree to which unstable superexcited neutral states contribute to the threshold ionization dynamics of liquid water has yet to be determined. Should indirect ionization channels – associated with initial photoexcitation or primary photoelectron impact excitation – be significant in the near-threshold ionization of liquid water, as in water's other phases, we expect such processes to contribute to the LET signal and potentially to the background electron distributions extending to the primary, directly-produced photoelectron signals.

Our approach to model the aforementioned scattering and potential indirect ionization mechanisms is very simple, using well-established scattering cross sections from gas-phase water and the available water photoabsorption and photoionization literature in an attempt to qualitatively explain our yet-to-be-detailed experimental observations.

**Correlation / anti-correlation of water photoelectron signal intensity with ionization and scattering cross sections: hν ≥ 20 eV**

We first determine the signal intensities of all water valence direct PE peaks and of the full (inelastic/indirect ionization) background signal (Fig. 1) based on Gaussian peak and baseline fits, as introduced above. The resulting direct PE peak fit areas are shown in Fig. 2A on the eKE x-scale associated with each direct PE peak (bottom axis), as well as on a photon-energy x-scale specifically pertaining to the water $1b_1$ ionization channel (top axis). The peak areas are presented as extracted from the fits, except for the $2a_1$ contribution which is scaled up by a factor of 7 to bring its values to approximate overlap with results for the other orbitals. This compensates the smaller $2a_1$ ionization cross section at the employed photon energies.[59, 60] Peak signal intensities are found to increase with decreasing eKE and then exhibit a steep decline to zero in the 5-15 eV eKE region; all orbitals exhibit the same trend fully in line with negligibly small beta-induced effects (as explained above). Here, a value of 'zero' means that the direct PE peak is so small or distorted that it can no longer be identified in the spectrum. Notably, the observed smooth signal variation up to the sudden drop scales approximately with the experimental partial photoionization CSs of the gas-phase water orbitals $1b_1$, $3a_1$, and $1b_2$, as concluded based on the matching eKE dependent photoionization CS curve (purple dashed line).[61] All three orbitals have very similar CSs; we detail how the displayed CS curve relates to the literature data in Fig. SI-3. Note that in an early LJ-PES publication[50] a rather large difference between the gas- and liquid-phase water ionization CSs was reported, especially for the $1b_1$ and $3a_1$ orbital ionization channels. However, these differences may have originated from a combination of overestimating the background using a simple Shirley-type subtraction procedure, uncompensated polarization-dependent intensity variations, and possible variations in the spectrometer transmission function, as discussed in the original publication.[50]

Before discussing the origin of the steep signal drop near 10-14 eV found in Fig. 2A, we additionally analyze the background-to-direct-PE signal ratio, displayed in Fig. 2B; here the background is the local background signal directly underneath the respective primary, directly-produced PE peak. Associated absolute background intensities are presented in Fig. SI-4. The same eKE axis is used as in Fig. 2A. Relative to the peak area, the underlying electron background signal increases in intensity towards low energies. Assuming the background signal originates from the same direct PE channels, we see that it solely scales with eKE but is independent of the ionized orbital. Hence, the background-to-direct-PE signal ratio is constant within our experimental error up until eKE ~13 eV. Notably, this is the same eKE region where the change in behavior is observed in Fig. 2A.

In order to explain the sudden changes of behavior observed in Figs. 2A and 2B, we consider the CSs for the various relevant inelastic electron scattering processes discussed in the literature as a function of eKE, which are shown in Fig. 2C. We deliberately choose the gas-phase CS values[25] for this comparison to emphasize that the effect is not exclusively a specific property of the liquid state, but mainly stems from the fact that the much higher density of liquid water leads to more pronounced scattering. Furthermore, the scattering CSs for aqueous solutions are admittedly less accurately known, with significant variance of the associated values being reported by different research groups.[56, 62, 63] This is partially due to the need to invoke additional processes, including several intermolecular vibrational energy transfer processes (over many water molecules) leading for instance to librations, translational displacement, and bending motion of the intermolecular hydrogen-bond coordinate,[64] all of which participate in the breaking and forming of hydrogen bonds. Rotational motion, on the other hand, is strongly suppressed by the hydrogen-bonding network. Notably, however, none of these details are directly considered in our qualitative signal analysis; we here focus on the cross-over from known electronic to alternative, *e.g.*, vibrational, scattering behaviors occurring in the 10-20 eV eKE region. Indeed, particularly the scattering processes occurring at low eKEs are the most difficult to model in liquid water, although it has been suggested[56] that these processes and their CSs are very similar to those associated with amorphous ice.[27]

An important inference from Fig. 2C is that the dominant processes occurring at higher eKEs are the electronic scattering channels (resulting in relatively high-energy losses)[25, 65-69] – *i.e.*, ionization (in blue) and excitation and dissociation (in green) – with the respective (gas-phase) CSs tending towards zero near 10-14 eV eKEs, when the eKEs approach the $VIE_{1b1}$ threshold and the electronic scattering channels begin to close. Importantly, the total ionization cross-section data shown in Fig. 2C integrates over direct and any indirect electron-producing channels. It should also be noted that, as compared to the gas phase, the (vertical) valence excitation and ionization energies approach each other in liquid water, with the respective CS curves expected to tend to somewhat different eKE values. In the liquid, the electronic excitation CS curves shift to higher eKEs by ~1 eV, with the direct ionization scattering channel CS curves tending to lower eKE values by about 1-2 eV.[50] This brings the onsets of the major electronic scattering processes in liquid water closer together and the cumulative electronic inelastic scattering CS towards the lower eKE region characterized by predominant vibrational scattering. Despite these shortcomings, the gas-phase CSs shown in Fig. 2C clearly indicate that upon decreasing the eKE below 10-14 eV, the CSs for vibrational scattering processes, partitioned here into vibrational stretch (in red) and vibrational bends (in orange),[25] quickly rise and assume similar values to those of the electronic processes in the case of KEs above the cross-over region. In the following, we will concentrate just on these known scattering processes in water, and later discuss the possibility of additional (so far unexplored) excitation and indirect ionization channels likely contributing in this energy region.

To our knowledge, no (non-vibrational) inelastic scattering processes have been reported to produce few-eV (*i.e.*, LET contributing) electrons following low-KE-electron (0-15 eV) collisions with either gas- or liquid-phase water. Dissociative electron attachment to neutral water molecules (processes of the type $e^- + H_2O \rightarrow OH + H^-$),[69, 70] although included in Fig. 2C, is of minor importance here as these excitation channels exclusively act as electron sinks. Furthermore, the respective (gas-phase) CSs (brown curve) are one order of magnitude smaller than their vibrational inelastic scattering counterparts in the 5-10 eV eKE region. The fact that vibrational CSs are so large at the low eKEs (and we do not even refer to the CS spikes in the KE < 2 eV range) would imply that a primary photoelectron produced in this range engages in many quasi-elastic (low-energy) scattering losses which gives rise to a broad signal approximately centered (with higher tendency towards lower eKEs) at the original peak position in the spectrum. In other words, the undisturbed photoelectron peak diminishes in height, since fewer electrons escape from the liquid with their nascent KEs. This is balanced by a build-up of a broad scattering background right underneath the same peak. It can therefore be argued that the observed discontinuities in Figs. 2A and 2B for eKE < ~13 eV thus reflect the transition from mainly electronic scattering channels to alternative quasi-elastic processes such as those arising through vibrational scattering. Also, dissociative electron attachment and potential indirect autoionization mechanisms may contribute to the diminished primary PE peak areas in the intermediate eKE region, and potentially close to the photoionization threshold. In more practical terms, both Figs. 1 and 2 are quantitative and illustrative demonstrations of, and actually handy reference data showing, the lowest photon energy at which any PE feature from liquid water can still be extracted (essentially) undisturbed by scattering effects. (In a related upcoming publication, we will discuss smaller additional primary PE peak distortions, specifically slight peak energy shifts, which already set in at eKEs below ~30 eV.)

In light of the low-eKE inelastic scattering processes discussed above, we consider the 20-eV photon energy spectrum shown in Fig. 1 in more detail. This corresponds to a <10 eV eKE of the water $1b_1$ peak, which is just below $VIE_{1b1}$. Based on Fig. 2C, the LET signal should entirely consist of IPED at a 20-eV photoexcitation energy. Surprisingly, however, this does not seem to be the case. A plot of the (integrated) LET signal intensity divided by the ionization channel-resolved direct valence PE peak areas, presented in Fig. SI-5, reveals a smooth decrease towards lower eKEs as secondary-electron-generating processes diminish. However, below eKEs of ~14 eV, a steep rise indicates that an additional high-CS and yet unknown low-KE electron generating process must contribute to the LET signal intensity, a point we will return to below. Nonetheless, we argue that IPED contributions remain large in these low primary eKE regions to some extent because the probability for direct secondary ionization is diminished. If primary electrons cannot engage in high-energy losses, which would essentially remove them from the measurable photoelectron signal (*i.e.*, the electron distribution which has sufficient energy to overcome the aqueous-vacuum surface barrier), they can travel much further in the liquid and undergo many quasi-elastic scattering events. In such a case, the liquid probing depth will significantly increase, *i.e.*, the liquid will become (more) 'transparent' for electrons with KEs below ~10 eV. For this reason, LET electron emission from semiconductors, including liquid water, is found to be up to an order of magnitude more intense as compared to that observed from metals. This effect depends on the material's bandgap, $E_{gap}$, and the electron yield has been found to be highest when $E_{gap}$ ~7 eV (compared to 8.9 eV for liquid water[19]).[71] Thus, the fraction of IPED *versus* SEED gradually increases as we change the photon energy from 60 to 20 eV (Fig. 1). However, as Fig. SI-5 implies, another type of scattering process, or perhaps an alternative primary but indirect few-eV electron-generating mechanism, must also be included to describe the LET signal change that occurs in the 10-14 eV eKE region. Our current understanding of electron-collision interactions in liquid water for eKE <14 eV seems to be insufficient, and very likely processes other than vibrational scattering need to be identified and considered in order to explain the observed large LET signal, and perhaps the suddenness of the behavioral changes highlighted in Figs. 2A and 2B.

**Liquid water photoemission spectra: towards hv ≤ $VIE_{1b1(l)}$**

Although eKEs of approximately 10 eV (referring to the bottom spectrum in Fig. 1) are already too small to support the extraction of undistorted nascent PE peaks, it is instructive to explore photoemission spectra measured at yet smaller photon energies, so as to approach conditions met in a number of previous laser-based experiments. We are particularly interested in cases where hν < 11.3 eV (*i.e.*, the $VIE_{1b1(l)}$). Results are shown in Fig. 3 for the ionization of a liquid-water jet using photon energies between 10-25 eV. As compared to the measurements leading to Fig. 1, a much smaller bias voltage (-4 V) has been applied, implying that the apparent water gas-phase contribution is not as effectively smeared out. Spectra are displayed such that the leading spectral features exhibit approximately the same peak height; the full intensity scale and presentation of approximate relative signal intensities is displayed in Fig. SI-6. Here, zero energy marks the position of $VIE_{1b1(l)}$ (=11.3 eV, with a 1.45 eV FWHM[45]) as measured in a high-KE experiment, exemplified by the 25-eV PE spectrum where $1b_1$ ionization still yields a clearly resolved (but already somewhat distorted) peak profile. All other spectra have been shifted analogously by hν - $VIE_{1b1(l)}$, so as to display all spectra on the same relative energy scale with respect to $VIE_{1b1(l)}$. Note that for hν ≤ 15 eV the full spectral ranges are captured, while for the 20- and 25-eV spectra $E_{cut}$ is off scale as is the IPED (and SEED) signal intensity. Spectra measured at the latter two energies were already presented in Fig. 1, but we now observe additional intensity in the -1 to -3 eV range (on our relative energy scale) arising from gaseous water, due to the smaller bias voltage.

By successively lowering the photon energy down to 10 eV, we can track the photoemission distribution resulting from ionizing the full liquid water $1b_1$ valence band down to ionizing just its lowest binding energy component. For example, the PE signal in the 10-eV spectrum of Fig. 3 contributes only to the very onset of the lowest energy, $1b_1$ ionizing transition (extending approximately from +1.5 eV to +3 eV on the relative scale). The energetic range over which $1b_1$ emission can occur increases with increasing photon energy, which is reflected in a wider spectral width, in all cases terminated by the IPED cutoff, as is best seen for the 10- to 14-eV spectra.

The observed transition from unresolvable to resolvable but distorted direct $1b_1$ peak profiles above 14-eV excitation energies in Fig. 3 is particularly notable. This threshold coincides with that observed on the eKE scale, *i.e.*, primary photoelectron impact, shown in Fig. 2, and strongly suggests that a common resonant behavior in liquid water both competes with direct photoionization (see Fig. 3) and effectively electronically inelastically scatters primary photoelectrons (see Fig. 2A) at excitation energies of ~14 eV, a point we will return to in a subsequent sub-section.

Considering Fig. 3 more generally, this data cumulatively shows that the water spectrum, *e.g.*, measured at hν = 10 or 11 eV, is mostly composed of IPED (and potential autoionization signals; possibly, now also the CS spikes for eKE < 2 eV, if existent in liquid water, may need to be considered) that may be mistaken for the direct $1b_1$ PE feature. Hence, analysis of aqueous solution PES experiments performed so close to the ionization threshold must carefully account for the prevalent fraction of IPED signal (as well as any indirectly produced photoelectrons) *versus* the residual direct, nascent PE signal if spectral misinterpretations are to be avoided.

**Photoemission spectra from 3 M NaCl aqueous solution close to a core-level ionization threshold**

With the results presented so far, several pertinent questions arise. Is the low eKE behavior described above unique to the valence ionization of liquid water, or is it also observed for aqueous solutes? Further, have the nascent low-kinetic-energy PE peaks vanished, or have they rather been 'hidden' underneath the intense background signal as the primary eKEs were reduced? A related question is whether there is some inherent experimental flaw, such as suppressed electron transmission and detection, when measuring low-energy electrons with a hemispherical electron analyzer, despite application of an electron accelerating bias voltage to the sample. To answer these questions and unequivocally demonstrate the universality of diminishing primary aqueous-phase PE feature intensities at kinetic energies below ~10 eV, we conducted electron time-of-flight (TOF) coincidence measurements from a 3 M NaCl aqueous solution liquid jet using soft X-ray photons. Here our focus is on the $Cl^-_{(aq)}$ $2p_{3/2}$ and $2p_{1/2}$ core-level PE spectra. The electron binding energies are 202.1 and 203.6 eV, respectively,[38] *i.e.*, much larger than those of the valence features considered above. Experimental results are presented in Fig. 4. Figure 4A shows the normal (non-coincidence) TOF photoemission spectra of the $Cl^-_{(aq)}$ 2p peak atop the LET spectral component, with the cutoff to the left. Note that electron counts are presented on a logarithmic scale to avoid cutting off the LET intensity for the higher-photon energy spectra. Upon lowering the photon energy towards the $Cl^-_{(aq)}$ 2p ionization threshold, the 2p peak moves to smaller eKEs, and when approaching ~10-14 eV, it starts to distort and broaden, and progressively decreases in intensity. This behavior is fully analogous to that shown for the liquid water valence ionization features in Fig. 1.

The LET signal in Fig. 4A mainly consists of inelastically scattered Auger electrons associated with the 2p core-hole decay (with some contribution from high-energy $Na^+$ and water direct PE electrons), and thus changes only slightly with photon energy. Subtraction of this background reveals the remaining $Cl^-_{(aq)}$ 2p PE signal, as plotted in Fig. 4B, with intensities now presented on a linear y-scale. Respective peak areas (extracted from a fit to the spin-orbit split 2p peak with two Gaussians) are plotted as a function of eKE, as shown by the black data points in Fig. 4C. The peak intensity is observed to rise as the eKE decreases from 18 eV to approximately ~12 eV, where the signal intensity suddenly drops towards zero intensity, seemingly highlighting the onset of quasi-elastic and/or resonant electronic scattering at this energy, fully analogous to the findings in Fig. 2A. Furthermore, as in Fig. 2A, the initial smooth signal rise follows the increase of the Cl 2p ionization CS with decreasing photon energy, depicted by the fitted CS curve (purple dashed) associated with the y-scale to the right (where the fit to the CS data is shown in detail in Fig. SI-7 of the SI).

To address the additional questions posed above regarding possible signal detection deficiencies, we also performed electron–electron coincidence measurements from the same solution to effectively reduce the inelastic scattering background. The results are presented in Fig. 4D and are obtained from a two-hit coincidence measurement and analysis, triggered by a fast Auger electron acquisition (150-200 eV) window. By selecting only two-hit events which include (mostly undisturbed) Auger electrons originating from the Cl 2p core-hole decay, the random inelastic background is suppressed by orders of magnitude (note the linear scale in panel D). Similar peak fits to those adopted with the non-coincidence data were applied to the coincidence data, however explicitly including the (unsubtracted) sloping background here (dashed black lines). The resulting eKE dependence of the peak areas is plotted in Fig. 4C (red points) for comparison with the non-coincidence measurement data. We observe the same behavior with eKE, a sudden drop in the direct PE peak area below an eKE ~ 12 eV. We note the similar eKE-dependent ionization trend to that observed with (nearly) neat water close to the valence ionization threshold (see Fig. 2A), and attribute the less abrupt decrease in intensity below eKE ~12 eV in the core-level spectra to the smaller photon energy steps used, 2 eV, in comparison to the liquid water valence data, 5 eV.

Collectively considering the aqueous solution data, we can confirm that the discussed inelastic scattering effects are not a property of the detection method. Furthermore, the observation of the same behavior for both the valence solvent and core-level solute features demonstrates that the primary photoelectron peaks are not just masked by a large inelastic scattering background, but really are diminished at low eKEs, as a general effect in photoemission from aqueous solutions. In the next sub-section, we build on these inferences and speculate on the role of indirect ionization processes in the near-threshold PE spectroscopy of liquid water, beyond the observed effects of vibrational inelastic scattering.

**Role of superexcited states and autoionization in near-ionization-threshold-excited liquid water**

Reviewing the observations reported above, we find that (1) the directly-produced primary $1b_1$ PE peak disappears at photoexcitation energies of $h\nu \leq 14$ eV (see Fig. 3), (2) a step-like decrease in the direct PE peak yields occurs for all direct valence ionization channels below a 10-14 eV eKE threshold (see Fig. 2A), (3) a step-like increase occurs in the LET signal as the direct valence PE feature eKEs drop below ~13 eV (see Fig. SI-5), and (4) similarly to that described in (2), a step-like decrease occurs in the direct PE signal from an aqueous solute (as opposed to the water solvent) at eKE values of ~12 eV, at photon energies ~190 eV beyond observations (1)-(3) (see Fig. 4D). These observations collectively and specifically identify a change in the ionization behavior of liquid water below an excitation energy threshold of 12-14 eV, with the same threshold being observed on two different excitation energy scales, *i.e.*, both the hν and eKE scales, respectively, associated with photon and electron impact excitation. While the change in the primary, direct PE signals on the eKE scale could be attributed to the transition from predominantly electronic inelastic scattering processes to vibrational alternatives (see Fig. 2C and the explanations above), the similar threshold observed on a hν scale cannot; vibrational inelastic scattering process are expected to dominate with similar CSs above and below hν ~ 12-14 eV (= eKE ~1-3 eV). This suggests that an alternative process is (at least partly) responsible for the loss of primary, direct PE peak intensity below the aforementioned thresholds. We propose that one or more resonant, neutral-state excitation routes exist in liquid water at energies between 10-14 eV and that they are (at least partly) responsible for the threshold behaviors. The associated, commonly populated superexcited state (or states) must be efficiently accessible both via photoexcitation and electron impact, where electric dipole selection rules will primarily govern the former. Furthermore, observation (3) indicates that the accessed superexcited states have a measurable autoionization yield and produce low-KE electrons.

Optically-bright, superexcited state resonances at 10-14 eV in liquid water would be expected to contribute to the associated photoabsorption spectrum. While broad absorption peaks are observed in the photoabsorption curve of hexagonal ice at 12.4 eV and 14.5 eV,[72] those features are found to be inhomogeneously broadened in liquid water and amorphous ice, resulting in a merged profile.[72, 73] The crystalline ice peaks and the broader amorphous ice structure have been respectively attributed to $4a_1 \leftarrow 3a_1$ and $4a_1 \leftarrow 1b_2$ valence-to-conduction-band (electric-dipole allowed) transitions, with such an assignment likely being extendable to liquid water. Concerning the fate of such superexcited states, autoionization is expected, as described in the following. A related higher-lying amorphous ice absorption feature centered at ~28 eV has been assigned to the $4a_1 \leftarrow 2a_1$ transition, with the resulting superexcited state found to decay through an autoionization process.[58] This process was observed to yield a broad secondary photoelectron spectrum peaking at an eKE of 11 eV and covering the 7-17 eV range, presumably via a $1b_2^{-1}$ indirect ionization process (where $VIE_{1b2(s)}$ is ~17.6 eV[74, 75]). We expect similar processes to occur in liquid water following 25-30 eV photon or primary electron impact excitation, where the latter may well be discernible in the hv > 50 eV data shown in Fig. 1; see the weak and broad secondary PE peaks imposed on the background signals at eKEs of 11-18 eV.

We now return to the four observations listed above. Both $4a_1 \leftarrow 3a_1$ and $4a_1 \leftarrow 1b_2$ photon and electron impact excitations of neutral liquid water are expected to occur between 10-15 eV. The resulting superexcited states are expected to decay through autoionization to produce the $1b_1^{-1}$ and perhaps $3a_1^{-1}$ (energetically accessible[50, 76]) cation states with low-KE electrons spanning 0-5 eV, via a similar mechanism as observed following $4a_1 \leftarrow 2a_1$ excitation in amorphous ice[58] (and potentially identified here in liquid water). Such $1b_1^{-1}$ and/or $3a_1^{-1}$ indirect autoionization processes would be consistent with observation (3), and in concert with vibrational inelastic scattering processes, may be responsible for observations (1), (2), and (4) as well. Notably, the plot in Fig. 2B, related to observation (2), might also be affected by the aforementioned $4a_1 \leftarrow 2a_1$ excitation and $1b_2^{-1}$ indirect autoionization process, given the 25-35 eV photoexcitation energies involved in the critical data plotted in Fig. 2B. Clearly, however, further investigations will be required to confirm and understand such near-threshold, indirect aqueous ionization processes in detail. In any regard, the inference that superexcited states and autoionization phenomena occur close to the ionization threshold in liquid water and are (at least partially) responsible for the loss of direct PE peak structures, further emphasizes the complexity underlying the low excitation energy PE spectra of liquid water. Furthermore, with regard to our division of the LET into SEED and IPED contributions, excitation-energy dependent indirect ionization channels seemingly need to be additionally included to fully describe our data and the overall LET signal produced at eKEs <15 eV, and perhaps <30 eV.

In the following we discuss one further potentially important inelastic scattering process and any effects it may have on our observations, namely the probability of photoelectrons emitted from the liquid jet colliding with and inelastically scattering from the surrounding gas-phase water molecules.

**Role of electron scattering with gas-phase water in a liquid-water-jet PES experiment**

Aside from laser-based studies, the majority of previous liquid-jet PES experiments were performed with photon energies considerably above associated ionization thresholds, in which case only electronic scattering is relevant, which does not impair our ability to detect unperturbed (nascent) PE peaks from the liquid phase. In this situation of sufficiently high-KE electrons, the contribution from the gas-phase signal is straight-forward to quantify. This is typically founded on a comparison of the spectrum measured from the liquid water jet and from the surrounding water vapor (the latter being selectively recorded by moving the jet out of the ionizing light beam focus); an example of which is presented in Fig. SI-1 (panel A, green curve *versus* panel B, red curve). Since the characteristic water orbital energies are different for the two phases, it is possible to even measure photoelectron angular distributions (PAD) for each phase.[62] This ability together with the fact that the PADs from the aqueous phase are distinctively different from the gas phase implies that the (Knudsen) gas-phase layer surrounding the jet has a negligibly small effect on the liquid-phase spectra. However, such a distinction of the gas-phase contribution is not as straight-forward at the low eKEs considered here; an issue that has been largely ignored so far. Why is this important? In the case of low-energy-loss channels, low eKEs, and when liquid and gas-phase water photoemission signal contributions are barely spectrally separated, an experimental determination of how much of the LET is due to electron-gas collisions is challenging. It is well-known, and a key aspect in the initial development of the liquid-microjet technology, that elevated gas pressures greatly diminish the PE signal.[77] With water, the situation is even more complicated as it does not exhibit a sharp boundary to vacuum, and rather the water density gradually decreases to the value of the gas on a length scale of about ~5 Å.[78, 79] This implies that in the case of the (low-photon energy and eKE) surface-sensitive PES experiments discussed here, the dense interfacial layer, with intermediate water density, must be inevitably considered as part of the liquid. What then remains to be explored is how much all other gas-phase water molecules on the way to the detector contribute to the LET and hence to the decrease of the initial intensity of the liquid-phase PE signal.

The gas density in the vicinity of the liquid jet quickly diminishes with distance as 1/r (where r is the radial distance from the nominal liquid-vapor interface boundary), until entering the skimmer orifice of the detection system where the pressure drops more rapidly (with $1/r^2$).[80] On this basis, a simple estimation (detailed in the SI) reveals that the effective thickness of the gas layer surrounding the liquid jet is too small for primary electrons of a few tens of eV KE to generate an appreciable LET signal. This conclusion is in accord with the fact that no LET signal is experimentally observed when probing only the gas layer around the liquid jet, as we demonstrate in Fig. SI-8 of the SI. We also recall that the dominance of quasi-elastic scattering translates into a considerable increase in electron escape depth from the solution, which becomes increasingly 'transparent' as the electrons have insufficient energy to electronically excite liquid water.

page 19

Possibly in such cases, the detection depth of the inelastically scattered electrons then approaches the optical penetration depth (inverse absorption coefficient) for UV excitation light, which is only on the order of ~20-60 nm in the 8-40 eV range, but rapidly increases by several orders of magnitude below ~7 eV.

## Conclusions

From liquid water and aqueous solutions alike, we observe both a rapid decrease in nascent, *i.e.*, undistorted, direct PE peak intensity and a rise of the background signal underlying the peaks when the photoelectron eKE falls below a critical energy of ~10-14 eV. This range coincides with the transition from known electronic to vibrational inelastic scattering channels, which vastly enhance quasi-elastic scattering and leads to deterioration of the nascent PE signal. Below the identified energetic threshold, PE features can no longer be reliably extracted (essentially) free from the effects of inelastic scattering, largely preventing the determination of correct VIEs and useful peak areas. Our results provide a reference eKE down to which PE features can still be extracted, largely undisturbed. This problem has only recently come under consideration in the aqueous phase, with sophisticated scattering models being developed with one aim being the retrieval of the nascent PE distribution. Yet, more knowledge of the underlying scattering process in liquid water and influencing factors is needed over an extended eKE range to refine the scattering models.

An additional important inference from this work is that following photon or electron impact excitation close to the ionization threshold of liquid water, indirect autoionization processes seemingly occur at the expense of direct photoemission, leading to effective production of threshold KE electrons, and a disproportionately large LET signal. We suggest that these low-KE electrons are produced via valence to $4a_1$ conduction band excitations, forming metastable superexcited states that subsequently autoionize. The specific processes occurring in the 10-14 eV photon or electron impact excitation range are thought to form internally excited $3a_1$ and $1b_1$ cation states and electrons in the 0-5 eV KE range, which undergo vibrational inelastic scattering prior to detection. Providing *direct* experimental evidence for these and other autoionizing superexcited valence states in liquid water represents an interesting and potentially important avenue of future research. This is particularly the case given that such processes are driven by some of the highest absorption CSs in liquid water and give rise to slow electrons, which are key contributors to radiobiological damage.[28, 29]

We emphasize that the underlying scattering phenomena discussed here may well be universal for solvents with similarly large band gaps as water, and moreover for all condensed matter exhibiting strong variation in scattering contributions as a function of eKE. Furthermore, one should keep in mind that quasi-elastic scattering is never completely turned off, and will give an additional error (albeit increasingly small in cases where higher eKEs are tended towards in liquid water) to any determined condensed-phase VIE or peak area. Very likely the scattering discussed here for macroscopic liquid water is different for water clusters and nano-droplets. Indeed, studies from large water clusters have not revealed the existence of a LET signal,[16] which is largely due to the fact that the clusters are significantly smaller in size (about 1 nm in size), resulting in PEs undergoing at most a single scattering event inside the clusters. Furthermore, there is an indication that the LET is significantly smaller in spectra from 100-nm nanodroplets.[81] It remains to be explored how the occurrence of the inelastic scattering background correlates with cluster/droplet size.

Finally, the results reported here imply that great care must be taken when analyzing aqueous-solution PES experiments performed close to a given ionization threshold, for instance when using multiphoton or pump-probe ionization schemes (for example with ~4.6 – 6.2 eV photons). In a related context, these considerations will be extended in a forthcoming publication to evaluate the reported and widely scattered values of the lowest VIE of liquid water,[45, 50, 82, 83] *i.e.*, the VIE attributed to the $1b_1$ molecular orbital, with particular attention to the KEs of the electrons detected and analyzed in different experiments. Based on the results presented here (as well as those to come), we make a specific recommendation for future liquid-phase PE spectroscopy measurements: where nascent eKEs, associated BEs, and PE peak profiles are to be accurately measured and reported, photon energies $\geq$ 30 eV above the ionization threshold of interest should be implemented.

## Acknowledgments

We would like to thank Ruth Signorell for many stimulating and fruitful discussions, as well as for the critical reading of and comments on this manuscript. S.T. acknowledges support from the JSPS KAKENHI Grant No. JP18K14178 and JSPS KAKENHI Grant No. JP20K15229. S.M., U.H., and B.W. acknowledge support by the Deutsche Forschungsgemeinschaft (Wi 1327/5-1). F.T., G.M., and B.W. acknowledge support by the MaxWater initiative of the Max-Planck-Gesellschaft. B.W. acknowledges funding from the European Research Council (ERC) under the Euopean Union's Horizon 2020 research and investigation programme (grant agreement No. 883759). D.M.N. and C.L. were supported by the Director, Office of Basic Energy Science, Chemical Sciences Division of the U.S. Department of Energy under Contract No. DE-AC02-05CH11231 and by the Alexander von Humboldt Foundation. We thank the Helmholtz-Zentrum Berlin für Materialien und Energie for allocation of synchrotron radiation beamtime at BESSY II. Some of the experiments were carried out with the approval of synchrotron SOLEIL (proposals numbers 99190181 and 20190130). We thank the technical service personnel of the SOLEIL chemistry laboratories for their helpful support.

# Figure Captions

**Figure 1:** Valence photoemission spectra from liquid water ionized at photon energies between 20 and 60 eV. Photoelectron peaks due to ionization of the water $1b_1$, $3a_1$, $1b_2$, and $2a_1$ orbitals are labeled. All peaks shift to higher kinetic energy (KE) with increasing photon energy according to KE = hv - BE. The signal associated with the inelastic scattering background is visible to the left of the water photoelectron peaks, and cumulates in the large scattering tail of the inelastic electron energy distribution (here denoted as the LET, low energy tail) curve close to zero KE. This LET curve is exposed by applying a negative bias voltage to the liquid sample, where the nominally applied voltage has been subtracted from the measured KEs to produce the KE scale shown here. Note that the measured cutoff signal will not necessarily coincide with zero KE after such a subtraction (as is the case here); the actual bias at the liquid jet is usually slightly different from that applied at the voltage source due to resistances and charge drops along the bias chain, and the potential presence of additional potentials. All spectra were fit by a series of Gaussians to account for the contributions of the nascent, undisturbed, direct PE peak signal (blue lines) and the inelastic scattering background signal (dashed black lines). See the main text for details.

**Figure 2:** Intensity of liquid water valence photoelectron peaks and the underlying inelastically scattered background, in comparison with ionization cross section data. Data are derived from the peak fits shown in Fig. 1. **A)** Peak areas of the $1b_1$ (red circles), $3a_1$ (green squares; sum of double-peak), $1b_2$ (blue triangles), and $2a_1$ (black diamonds) direct photoelectron features *versus* electron kinetic energy. The gas-phase ionization cross section, averaged over the $1b_1$, $3a_1$, and $1b_2$ molecular orbitals,[61] is overlaid onto the data as a purple dashed line, and is referred to the y-scale on the right (see text and Fig. SI-3 in the SI). Error bars depict the breadth of results obtained from running a least-squares fit of the spectra with varying model parameters and constraints. The top axis shows the photon energy specifically corresponding to $1b_1$ orbital ionization channel. **B)** Inelastically scattered local-background-signal strength at the respective peak position, relative to the peak areas of the signal components shown in A; the background (dashed lines in Fig. 1) was integrated over range of each peak's FWHM. The scaling factors the $1b_1$, $3a_1$ and $2a_1$ account for the fact that each peak sits atop a different background because of its relative position in the spectrum, which may include scattering contribution from higher eKEs (see also Fig. SI-4). The ratio of local background signal height *versus* nascent peak area rises steeply below ~13 eV KE. **C)** Cross sections of various electron scattering channels for the (gas-phase) $H_2O$ molecule from Refs. [25, 69, 70]. Ionization (blue), vibrational stretch (red), and vibrational bend (yellow) from Ref. [25]; direct dissociation following excitation (green) from Ref. [69]; dissociative electron attachment (brown) from Ref. [70].

**Figure 3:** Photoemission spectra of liquid water obtained for photon energies of 10-25 eV, which covers energies above and below $VIE_{1b1(l)}$ (= 11.3 eV). In the absence of a calibrated transmission function of the electron analyzer, spectra measured at hv = 10-14 eV are scaled to yield the same peak heights. Spectra measured at hv = 15-25 eV are scaled such that the signal height of the $1b_1$ photoelectron peak has approximately the same height as the features of the low-photon energy spectra; this is a convenient procedure, not based on scientific grounds, but sufficient for the present purpose. All spectra are shifted so as to compensate the difference in photon energy according to hv - $VIE_{1b1(l)}$ ($VIE_{1b1(l)}$ = 11.3 eV). In this presentation it can be immediately seen that the 10-14 eV spectra are composed of a prevailing LET contribution; the true photoemission signal associated with ionization of the water $1b_1$ orbital cannot be identified. Note that the apparent intensity difference between the 14- and 15-eV spectra is a result of the aforementioned intensity scaling procedure. The as-measured spectra (before normalization) are shown in Fig. SI-6. The 15-eV spectrum does exhibit a small $1b_1$ shoulder, and this spectrum can hence be displayed in the same manner as the 20- and 25-eV spectra. On the other hand, no $1b_1$ PE signal can be identified in the 14-eV photoemission spectrum, and this spectrum is scaled in the same way as all the spectra measured at yet lower photon energies.

**Figure 4:** Time-of-flight-based PE spectra of the Cl 2p doublet peak from 3 M NaCl aqueous solution at various photon energies close to the ionization edge. **A)** Non-coincident electron spectrum (log scale) after time-to-energy conversion. The photon energy was successively lowered to bring the peak's eKE towards and below the critical ~10-14 eV region. The PE peak intensity successively diminishes and eventually almost disappears below ~10 eV as the nascent direct PE signal is degraded by scattering. **B)** The major inelastic scattering background from liquid water is subtracted, which reveals the leftover Cl 2p signals. Although a peak is still visible at lower eKE, inelastic scattering leads to a broadened and asymmetric shape. **C)** Peak areas as extracted from the spectra in panels B and D. A decline of peak intensity below ~12 eV is observed, coinciding with the crossing from electronic to vibrational, *i.e.*, quasi-elastic, scattering (compare to Fig. 2C). The Cl 2p photoionization CS is shown for comparison (see text and Fig. SI-7 in the SI). Again, error bars depict the breadth of results obtained from running a least-squares fit of the spectra with varying model parameters and constraints. **D)** Cl 2p PE signal of the slow electrons extracted from a two-hit coincidence analysis of the same measurement and data shown in panel A. Here, the inelastic background signal is vastly reduced. Even when suppressing the background signal through coincidence analysis, the nascent 2p feature spectral profiles cannot be retrieved at low eKE values. More specifically, this demonstrates that below ~10 eV KE, the PE signals from solutes (which are not strongly surface-active) are distorted and predominantly suppressed.

# Figures

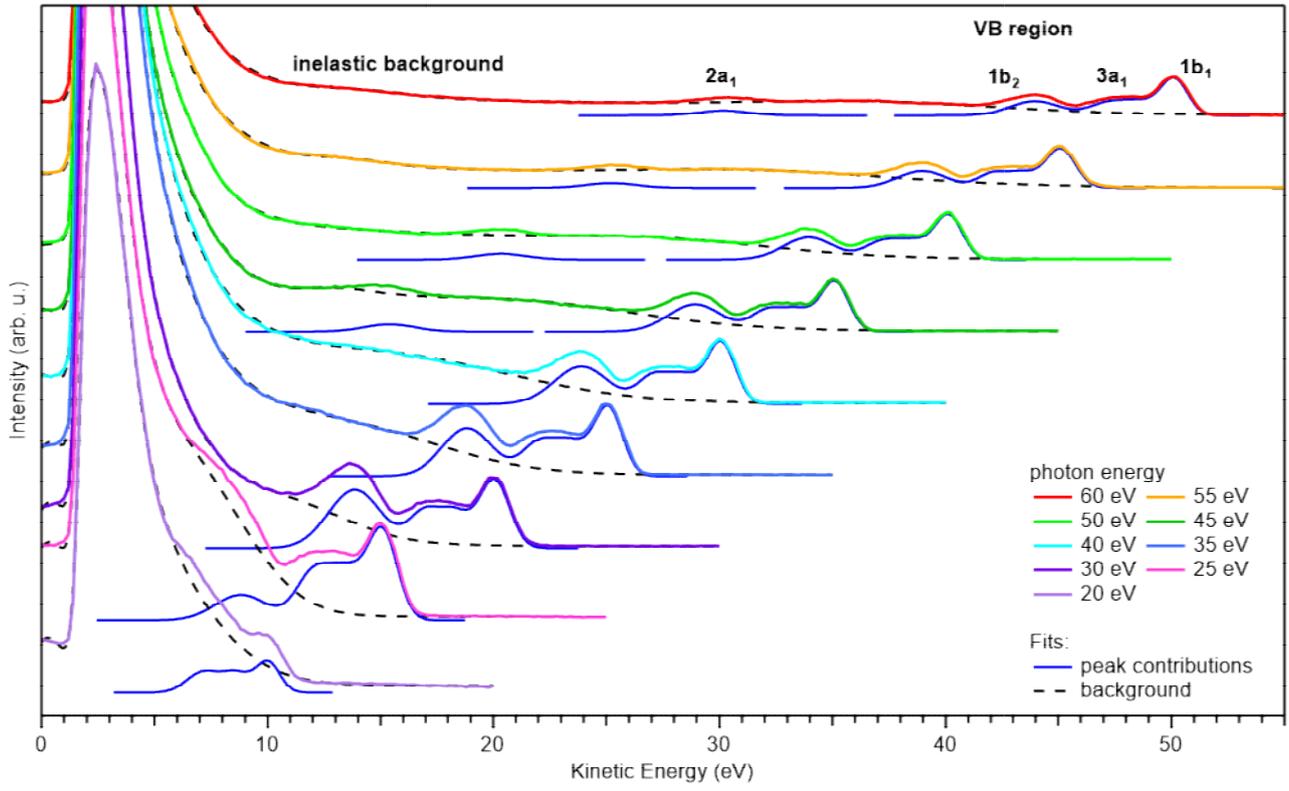

**Figure 1**

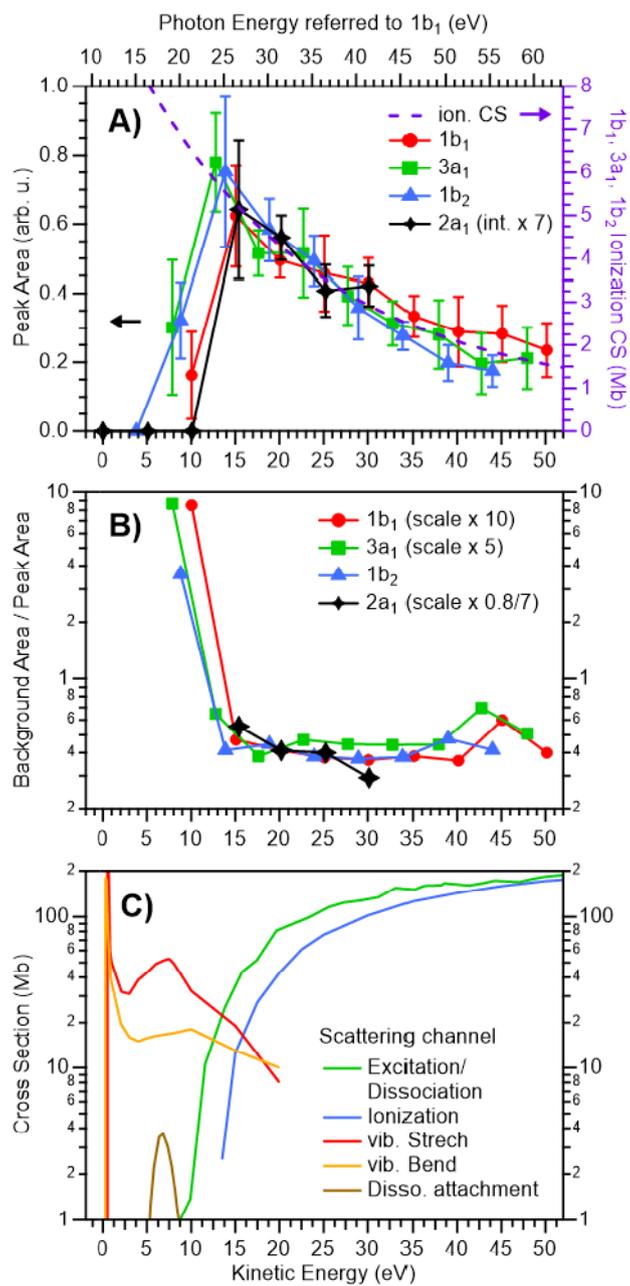

**Figure 2**

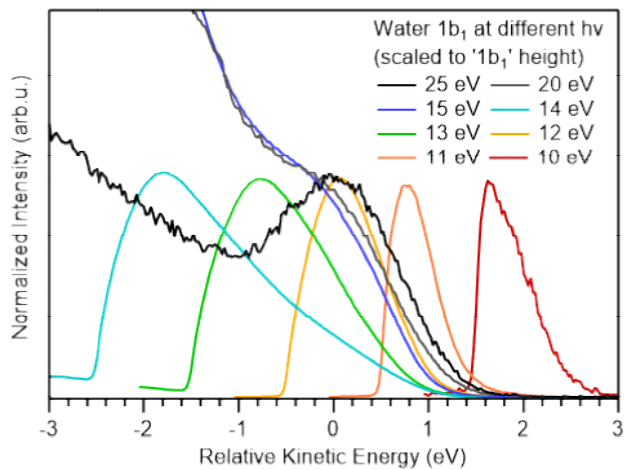

**Figure 3**

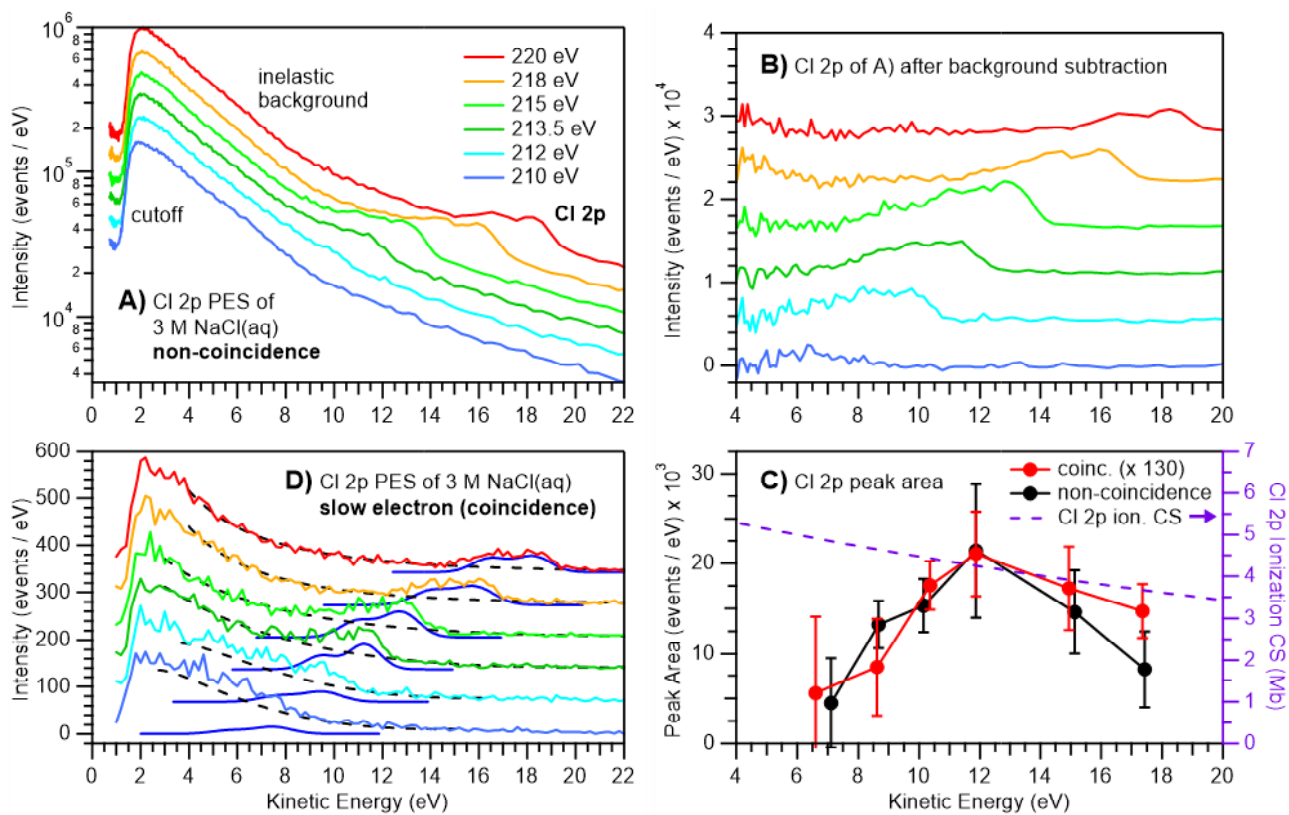

**Figure 4**

# Notes and References

†. Photoemission includes direct photoelectrons and any electrons emitted by some second-order process, e.g., Auger decay.

‡. Technically, recent laser developments can bridge this energy gap, with high harmonic generation sources routinely providing 100 eV photon energies, and even extending beyond 300 eV in a few laboratories, albeit with additional experimental complexity. See, e.g., Ref. [84].

# Supplementary Information

# Low-energy constraints on photoelectron spectra measured from liquid water and aqueous solutions

Sebastian Malerz[1], Florian Trinter[1,2], Uwe Hergenhahn[1,3], Aaron Ghrist[1,4], Hebatallah Ali[1,5], Christophe Nicolas[6], Clara-Magdalena Saak[7], Clemens Richter[1,3], Sebastian Hartweg[6], Laurent Nahon[6], Chin Lee[1,8,10], Claudia Goy[9], Daniel M. Neumark[8,10], Gerard Meijer[1], Iain Wilkinson[11]*, Bernd Winter[1]*, and Stephan Thürmer[12]*

[1] *Molecular Physics Department, Fritz-Haber-Institut der Max-Planck-Gesellschaft, Faradayweg 4-6, 14195 Berlin, Germany*
[2] *Institut für Kernphysik, Goethe-Universität, Max-von-Laue-Strasse 1, 60438 Frankfurt am Main, Germany*
[3] *Leibniz Institute of Surface Engineering (IOM), Department of Functional Surfaces, 04318 Leipzig, Germany*
[4] *Department of Chemistry, University of Southern California, Los Angeles, CA 90089, USA*
[5] *Physics Department, Women Faculty for Art, Science and Education, Ain Shams University, Heliopolis, 11757 Cairo, Egypt*
[6] *Synchrotron SOLEIL, L'Orme des Merisiers, St. Aubin, BP 48, 91192 Gif sur Yvette, France*
[7] *Department of Physics and Astronomy, Uppsala University, Box 516, SE-751 20 Uppsala, Sweden*
[8] *Department of Chemistry, University of California, Berkeley, CA 94720 USA*
[9] *Centre for Molecular Water Science (CMWS), Photon Science, Deutsches Elektronen-Synchrotron (DESY), Notkestraße 85, 22607 Hamburg, Germany*
[10] *Chemical Sciences Division, Lawrence Berkeley National Laboratory, Berkeley, CA 94720, USA*
[11] *Department of Locally-Sensitive & Time-Resolved Spectroscopy, Helmholtz-Zentrum Berlin für Materialien und Energie, Hahn-Meitner-Platz 1, 14109 Berlin, Germany*
[12] *Department of Chemistry, Graduate School of Science, Kyoto University, Kitashirakawa-Oiwakecho, Sakyo-Ku, Kyoto 606-8502, Japan*

# Figures

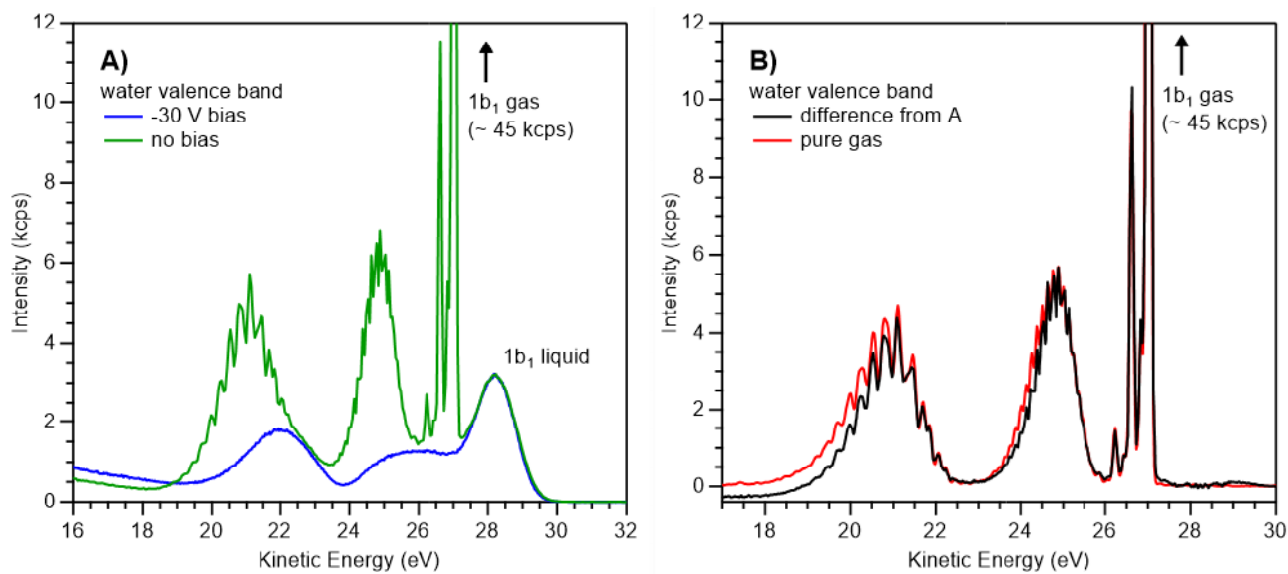

**Figure SI-1: A)** An exemplary photoelectron (PE) spectrum of the liquid water valence band measured with a He II α emission light source (hν = 40.813 eV). In the case of the blue spectrum, a -30 V bias voltage was applied to the jet, while the green spectrum was measured with a grounded jet. The kinetic energy offset imposed by the bias voltage has been corrected for and the blue spectrum was scaled to yield the same height for the liquid $1b_1$ peak. With the bias applied, the spectrum is almost completely free of gas-phase signal contribution which gets smeared out to lower kinetic energies (before correction for the effect of the bias voltage). The somewhat larger intensity of the blue spectrum for eKE < 20 eV is a consequence of small analyzer transmission changes when measurements are made at the 30 eV higher kinetic energy (KE). **B)** The difference spectrum between the green and blue curves in panel A is shown in black, and a spectrum measured from pure water vapor is shown in red for comparison. The latter spectrum was scaled to the same height and convolved with a Gaussian (width = 0.05 eV) to match the limited experimental resolution of the black difference spectrum.

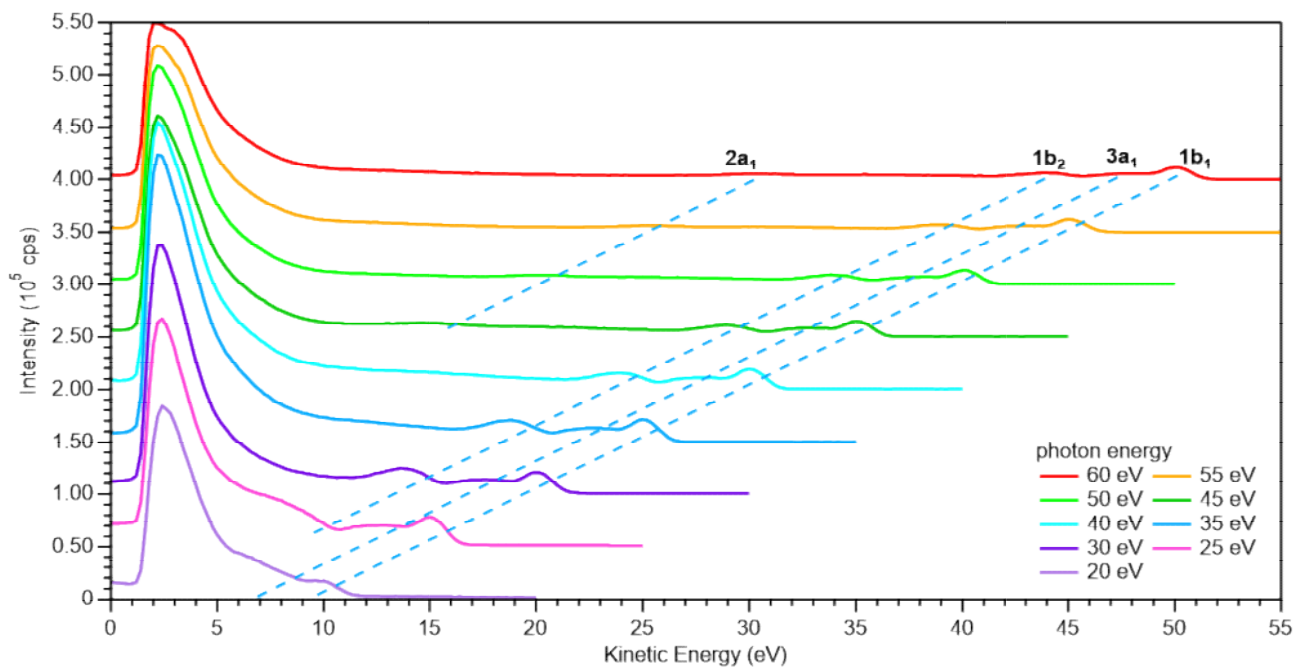

**Figure SI-2:** The same spectra as shown in Fig. 1 of the main text, but scaled to display the full magnitude of the low-energy tail (LET) curves. Each successive spectrum shown here is offset on the y-axis by 50,000 counts per second (cps).

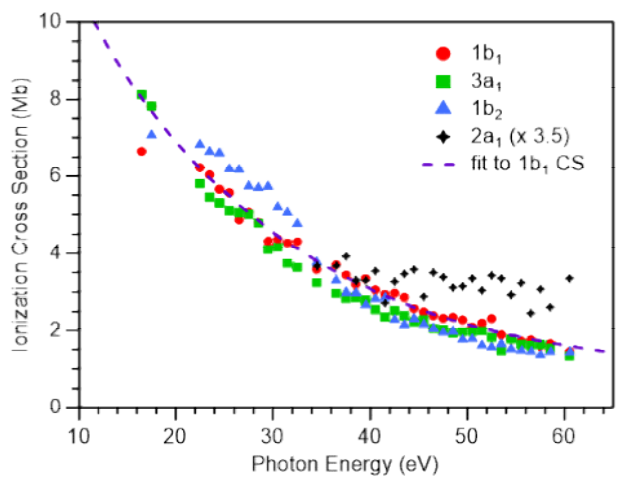

**Figure SI-3:** Experimental gas-phase photoionization cross sections (CSs) of the $1b_1$ (red circles), $3a_1$ (green squares), $1b_2$ (blue triangles), and $2a_1$ (black diamonds) orbitals from Ref. [1]. The purple dashed line represents an exponential fit of the $1b_1$ CS data; this approximate representation of the photoionization CS of the $1b_1$, $3a_1$, and $1b_2$ orbitals is used for comparison in Fig. 2A in the main text. The data we show here were obtained by electron impact at low momentum transfer, as this is the only available data set covering the threshold region. Photoelectron measurements, available above 30 eV, agree with the former within the scatter of the data.[2]

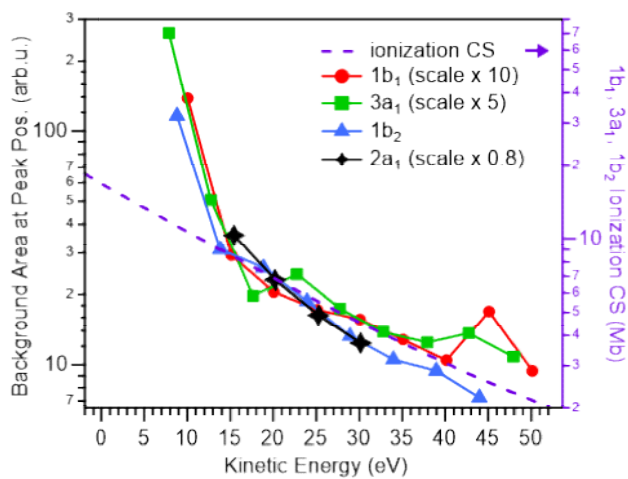

**Figure SI-4:** Absolute inelastic background signal strength at the respective peak position before normalization to the peak areas (which is shown in Fig. 2B in the main text). Background intensities were extracted at the $1b_1$ (red circles), $3a_1$ (green squares), $1b_2$ (blue triangles), and $2a_1$ (black diamonds) peak positions by integrating over the FWHM of each peak. The signal scaling accounts for the fact that the background below the $2a_1$ ($1b_1$) peak is relatively high (low) due to the peak's relative position in the spectrum. A representation of the ionization cross sections for the $1b_1$, $3a_1$, and $1b_2$ ionization channels from Ref.[1] is overlaid onto the data as a purple dashed line using the scale to the right (see main text and Fig. SI-3). The background smoothly increases proportionally to the CS towards lower eKE until the 10-15 eV region is reached, where it starts to deviate from this trend and rises towards higher intensities.

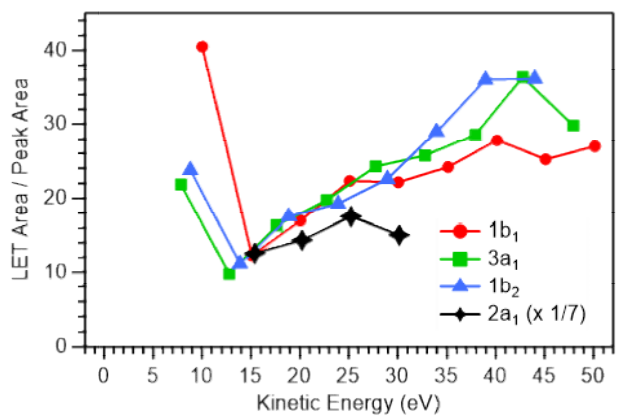

**Figure SI-5:** Low-electron-kinetic-energy tail, LET, area vs. peak area. The LET was integrated from 1-5 eV after correction for the residual gas-phase contribution and then normalized to (divided by) the $1b_1$ (red circles), $3a_1$ (green squares), $1b_2$ (blue triangles), and $2a_1$ (black diamonds) peak areas.

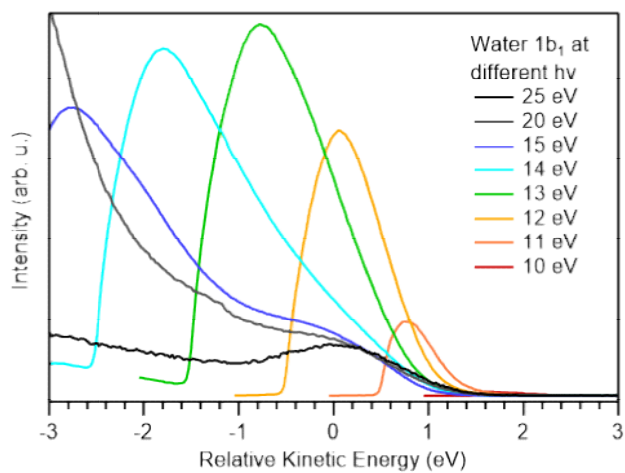

**Figure SI-6:** Photoemission spectra from liquid water obtained for photon energies of 10-25 eV, which covers energies above and below $VIE_{1b1(l)}$ (= 11.3 eV); the figure is based on the same data as Fig. 3 in the main text. As before, all spectra are shifted so as to compensate for the difference in photon energy according to hv - $VIE_{1b1(l)}$ ($VIE_{1b1(l)}$ = 11.3 eV). Unlike in Fig. 3, LET intensities are not clipped but rather displayed to yield the approximate relative intensities. For that, spectral intensities have been adjusted for photon flux and for (very approximate) changes in analyzer transmission; the latter is found to be sufficient for our qualitative analysis.

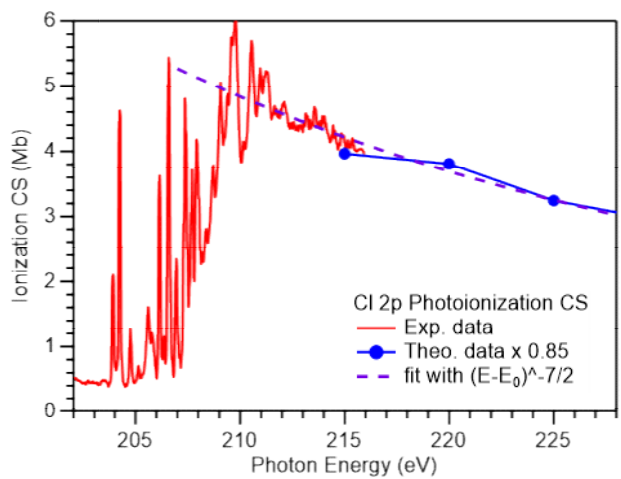

**Figure SI-7:** 2p photoionization cross sections (CSs) of atomic Cl: Experimental data from Ref. [3] and theoretical data from Ref. [4] scaled by a factor of 0.85. The purple dashed line represents a fit with $(E-E_0)^{-7/2}$ of the experimental CS from 209 eV to 216 eV; this approximate representation of the photoionization CS is used for comparison in Fig. 4D in the main text.

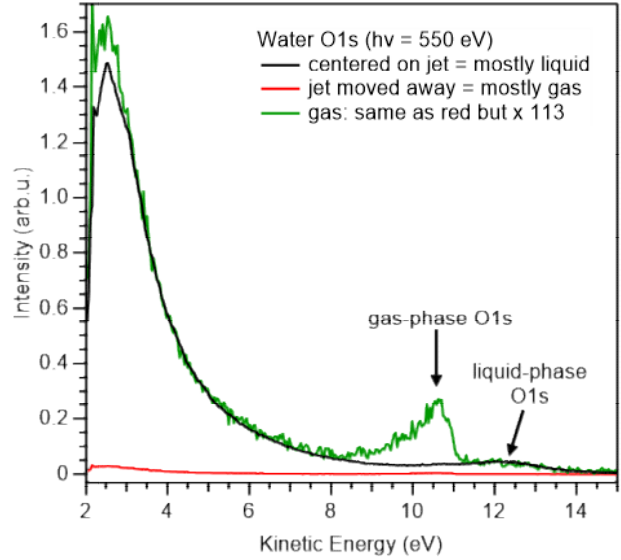

**Figure SI-8:** Photoemission spectra from liquid and gas-phase water measured at 550 eV photon energy using TOF spectroscopy, exhibiting the respective O 1s photoelectron peaks and LETs. The black curve shows the signal when the overlap between the liquid jet and the X-ray focus is optimized. The red curve was measured when the jet position was shifted sufficiently so that the X-ray spot has almost no overlap with the liquid jet, resulting in almost exclusive ionization of the surrounding gas-phase molecules. The green curve shows the red curve multiplied by a factor 113 which yields matching LET spectra. We find that the small contribution from the liquid phase essentially yields a scaled-down liquid water photoemission spectrum, implying that there is no LET signal generated in the gas phase. Spectra were measured with the magnetic bottle set-up described in conjunction with Fig. 4 of the main text. The asymmetry of the O 1s gas-phase peak results from both vibrational excitations[5] and asymmetric peak broadening due to post-collision interactions.

*Estimate of the number of electron - gas-phase water collisions in a liquid-jet PES experiment*

In the following we estimate the importance of collisions between electrons emitted from the liquid jet with molecules in the gaseous water surrounding the liquid jet. We assume a sharp gas-liquid interface and a vapor pressure above the liquid corresponding to the equilibrium water vapor pressure of 8 mbar at 4 °C. Furthermore, the gas-phase pressure is taken to drop linearly with distance, $r$, in the radial direction from the water-jet surface. The distance between the jet and the skimmer at the entrance of the electron spectrometer is taken as $r_1 = 0.5$ mm, and collisions after electrons have passed the differentially pumped skimmer aperture are considered unimportant. We can then estimate an effective vapor column $\mu$ (with the dimension of (particle number density) * length = inverse area, i.e., cm$^{-2}$) which characterizes the integrated gas density the electrons have to pass before being detected, according to Ref.[6]:

$$\mu(r_0, r_1) = \int_{r_0}^{r_1} dr\, n_0 \frac{r_0}{r} = n_0 r_0 \ln\left(\frac{r_1}{r_0}\right) \qquad (1).$$

Here, $r_0$ is the jet radius and $n_0$ is the particle number density at the liquid-vapor interface. Inserting $r_0 = 12.5$ μm and $r_1 = 500$ μm, we arrive at $\mu \sim 1 \times 10^{15}$ cm$^{-2}$. Using the Lambert-Beer law in the following form to determine the intensity reduction of a beam passing some medium:

$$I = I_0 \exp(-nl\,\sigma) \qquad (2),$$

with $I_0$ and $I$ describing the original and the reduced intensity, $n$ and $l$ the density and path length through the medium, and $\sigma$ the cross section for some scattering process, we can insert our estimated value ($\mu \sim 10^{15}$ cm$^{-2}$) of the effective vapor column in place of $nl$. Taking for example the cross section of some inelastic process as $10^{-16}$ cm$^2$ (100 Mb), we then calculate a 90% (exp(-0.1) $\cong$ 0.9) transmission of the electron beam. Thus, the sudden drop in liquid-phase peak intensity in the 10-13 eV region cannot be explained by scattering in the gas phase, especially since the electronic scattering channels responsible for reducing the peak intensity decrease in this electron kinetic energy region.

It is interesting to note that the exponential term in Eq. (2), $\exp(-nl\,\sigma) = \exp(-0.1)$, can be interpreted as the $P(k=0)$ value of the Poissonian probability distribution for the number of collisions, $k$, encountered by an electron on its path through the gas phase, written as:

$$P_\lambda(k) = \frac{\lambda^k}{k!}\exp(-\lambda) \qquad (3).$$

Here, $\lambda$ is the probability parameter. The argument of the exponential function thus corresponds to the average number of collisions encountered due to the process with cross section $\sigma$. We have 0.1 collisions for a $\sigma$ of $10^{-16}$ cm$^2$. In view of the value of $\mu$ estimated above, it is seen that only processes with cross sections exceeding 1000 Mb can lead to an appreciable number of collisions, under the assumptions given above. Referring to the rich literature on electron scattering on gas-phase H$_2$O, we find that only elastic or quasi-elastic scattering involving molecular rotation may reach cross sections in that range. The latter scattering processes may influence the angular distribution function, but will not lead to significant changes in the kinetic energy of the electrons. Thus, for primary electrons of several tens of eV, the contribution to the inelastic background and LET must be small. Indeed, we do not find any contribution of the gas-phase water molecules to the measured LET, as demonstrated in Fig. SI-8.